\documentclass[11pt]{article}
\usepackage{moriond}
\usepackage{epsfig,hyperref}

\newlength{\ziffer}

\newcommand{\TeV}{\,\mbox{Te\kern-0.2exV}}
\newcommand{\GeV}{\,\mbox{Ge\kern-0.2exV}}
\newcommand{\mGeV}{\,\mathrm{Ge\kern-0.2exV}}
\newcommand{\MeV}{\,\mbox{Me\kern-0.2exV}}
\newcommand{\keV}{\,\mbox{ke\kern-0.2exV}}
\newcommand{\eV}{\,\mbox{e\kern-0.2exV}}

\newcommand{\bea}{\pagebreak[3]\begin{samepage}\begin{eqnarray}}
\newcommand{\eea}{\end{eqnarray}\end{samepage}\pagebreak[3]}
\newcommand{\beq}{\begin{equation}}
\newcommand{\eeq}{\end{equation}}

\newcommand{\as}{$\alpha_s$}

\newcommand{\asb}{$\alpha_0$}

\newcommand{\abb}{Fig.~\ref}
\newcommand{\fig}{\abb}

\newcommand{\eq}[1]{Eq.~(\ref{#1})}

\newcommand{\oas}{$\cal O$($\alpha_s^2$)}
\newcommand{\oass}{$\cal O$($\alpha_s^3$)}

\newcommand{\PDG}{EurPhysJC15_1}
\newcommand{\siggi}{Abreu:2000ck}
\newcommand{\lepqcdwg}{QCDWG_LEPFest}
\newcommand{\lepewwg}{LEPEWWG}
\newcommand{\rhadBethke}{Bethke:2000ai}

\begin{document}
\vspace*{4cm}
\title{Hadronic event structure, power corrections 
       and the strong coupling at LEP}

\author{ Daniel Wicke  }

\address{Bergische Universit\"at-GH, Gau{\ss}str. 20, D-42097 Wuppertal,
  Germany\\ {\tt Daniel.Wicke@cern.ch}}

\maketitle\abstracts{
Infrared and collinear events shapes are suited to directly probe properties 
of hard QCD. They are traditionally used to measure the strong coupling and to
test the gauge structure of QCD. 
Perturbative predictions exist in several variations all of which depend on
the renormalisation scheme leading to large theoretical uncertainties in the 
determination of \as.  
%In order to match perturbative predictions
%with data the non-perturbative effects of hadronisation have to be taken into
%account. Beside MC models the analytical  power corrections are now
%widely used.
To overcome this dominating error more and more schemes for setting the
renormalisation scale are investigated. The application of RGI perturbation
theory shows an incredible 
small spread of $\alpha_s$ indicating a reduced uncertainty and allows a
measurement of the $\beta$-function directly from mean values.
}

\section{Introduction}
Event shapes 
are sensitive to the strong coupling, \as, and the gauge structure of the
strong force.
However, the observables investigated (mean values, higher moments
and normalised distributions) don't depend on the production rate nor on
the event orientation and are therefore independent of the electroweak
production process. They are thus directly connected to fundamental properties
of the strong force.

QCD predictions for these observables 
exist in NLO and NLLA, but their 
combination  suffers from ambiguities in avoiding
double counting. Of the available matching
schemes  $\log R$ is the most popular.
All calculations depend on the unphysical renormalisation scale, 
which is varied in order to estimate theoretical uncertainties.

The conversion of perturbatively accessible partons into hadrons may have a
significant impact on the final value of an event shape observable.
In order to match perturbative predictions
with data these non-perturbative hadronisation effects 
thus have to be taken into
account.
Traditionally the only way to correct for these non-perturbative effects
was the application of Monte Carlo models, which
suffer from a large number of free parameters that need to be tuned.
Since a few years the analytical ansatz of power corrections~\cite{PhysLettB352_451,NuclPhysB511_396}
with only one
free parameter is used as an alternative.

\section{Measurements of the strong coupling}

\begin{figure}
\begin{center}
\vspace*{-1.4cm}~\\
\epsfig{file=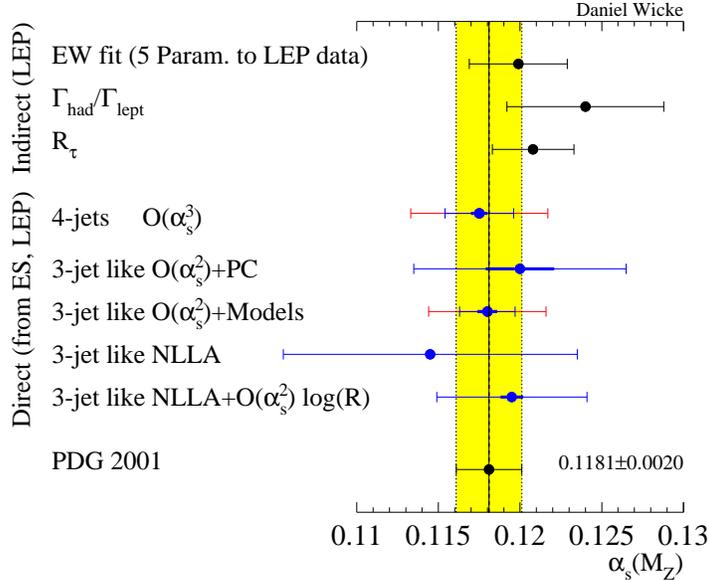,width=0.6\textwidth}\vspace*{-5mm}
\end{center}
\caption{\label{alphasbar}%
Comparison of
direct measurements of the strong coupling from event shapes at LEP
with indirect results from LEP and the current world average.
%The indirect results\cite{\lepewwg,\rhadBethke,\PDG}
%assume the correctness of electroweak sector of
%the standard model.
%\oass \cite{DELPHI2001-059_CONF487},
%\oas with power corrections \cite{Salam:2001bd},
%\oas \cite{\siggi},
%NLLA \cite{Akers:1995rx,\siggi},
%\oas+NLLA\cite{\lepqcdwg} with MC models.
}
\end{figure}

To give an overview \fig{alphasbar}
compares  measurements of the strong coupling from several
methods used at LEP to the current PDG world average~\cite{\PDG}.

Beside the results from the LEP QCD working group~\cite{\lepqcdwg},
a combination of pure NLLA results~\cite{Akers:1995rx,\siggi},
an application of \oas\ and \oass\ with MC models and optimised
scales~\cite{\siggi,DELPHI2001-059_CONF487} and results from \oas\ with power
corrections~\cite{Salam:2001bd} are shown.
The error contribution from the renormalisation scale was
adapted to use a consistent
variation of $\mu/Q$ by a factor 2 around the central value. 
The increase, if any, 
is shown as extra (red) error bars.

These direct methods show a good consistency within the quoted errors which
are in all cases dominated by the contribution from the renormalisation scale 
variation.

The indirect determinations of the strong coupling stem from the investigation
of $\tau$ hadronic branching ratio $R_\tau$~\cite{\PDG},
from the hadronic width of the $Z$~\cite{\rhadBethke}
and from a five parameter fit to electroweak precision data~\cite{\lepewwg}.
They show a good consistency with the direct results  and with
the world average within the given errors.

\section{Mass effects in event shape observables}
In order to improve the consistency in the description of event shape
observables it is important to take mass effects into account.
Mass effects arise from heavy quarks as well as from non-zero hadron masses.
The size of these effects depend on the observable and the energy.

Depending on the observable mass effects from stable
hadrons enter either indirectly
via energy-momentum-conservation (e.g. for thrust) or directly into the
observable (e.g. jet masses). The direct dependence on the hadron masses can
be avoided by an appropriate redefinition of the four-momenta used to
calculate the observable: 
$  p = (\vec{p},E)  \longrightarrow (\vec{p},\left|\vec{p}\right|)$
({$p$}-scheme) or
$  p = (\vec{p},E)  \longrightarrow (\hat{p}E,E)$ ({$E$}-scheme).

\begin{figure}[tb]
  \begin{center}
\hfill\epsfig{file=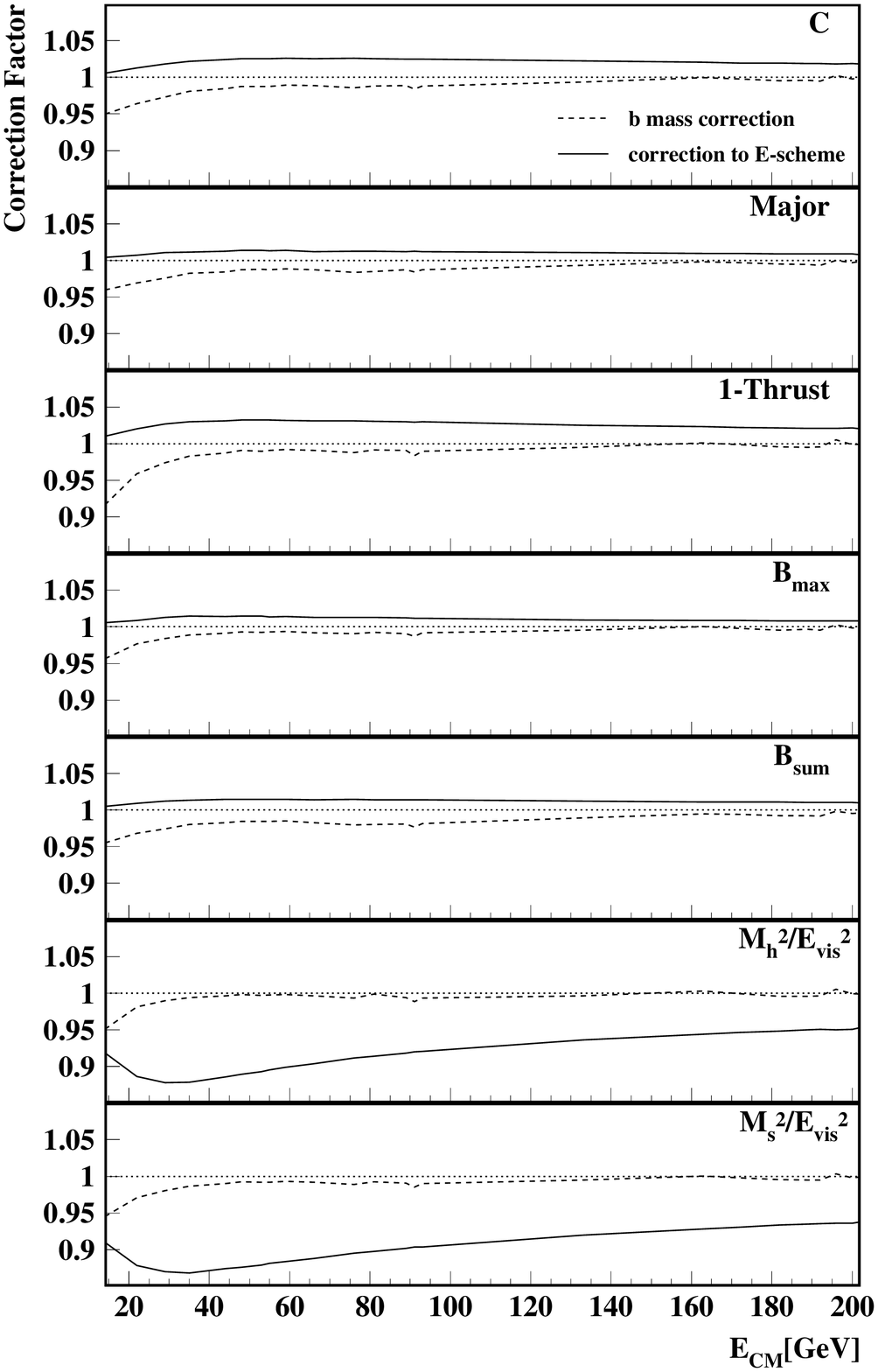,width=0.42\textwidth}\hfill
\epsfig{file=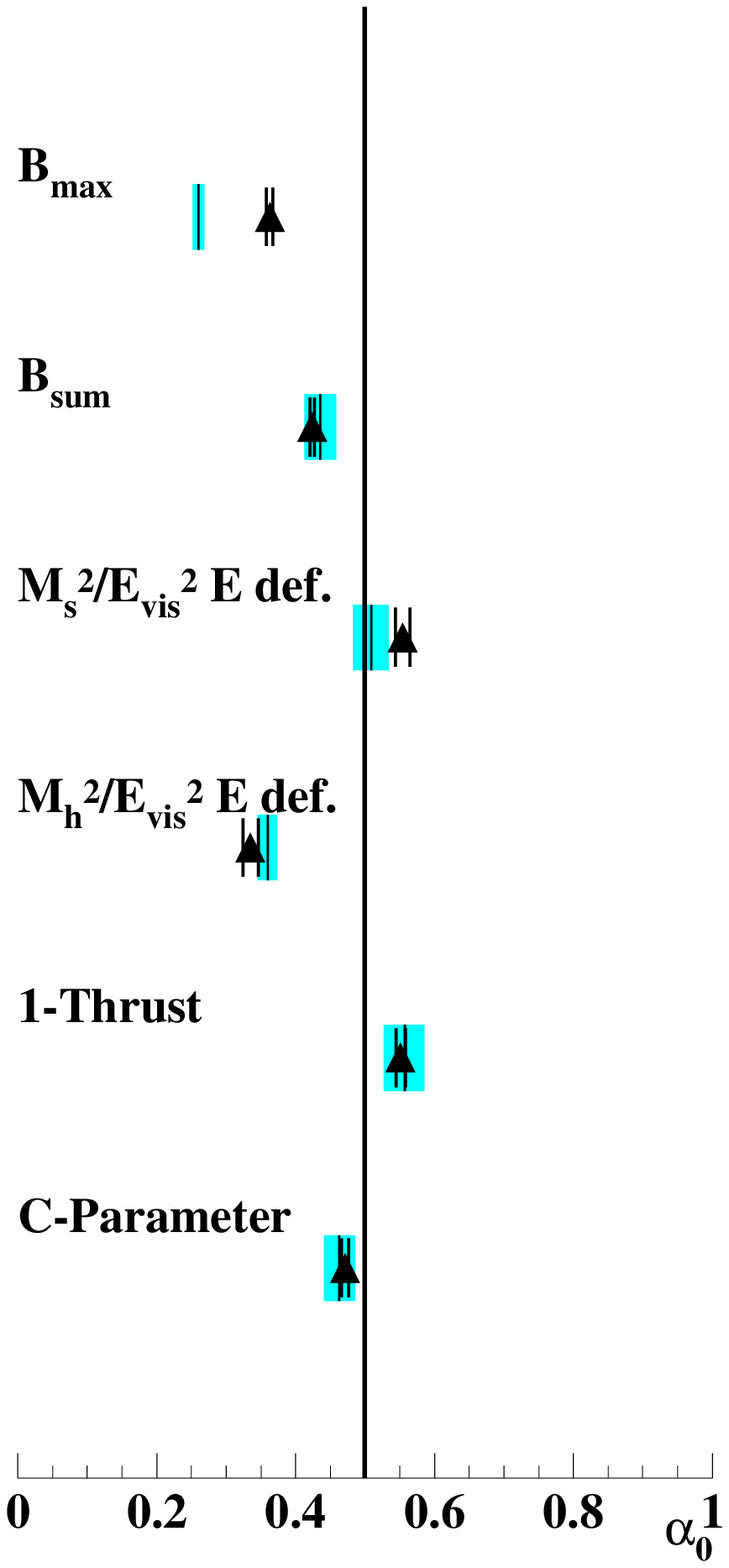,width=0.3\textwidth}\hfill\null
  \end{center}
\caption{\label{fig:a0prediction}\label{fig:masseffects}%
Left: Size of mass corrections to event shape means. Full line shows
difference between standard and $E$-scheme definition indicating hadron mass
effects. Dashed line shows influence of $B$-hadrons.
Right: 'Prediction' of the power correction parameter \asb\ from RGI (blue band) 
compared to results from two-parameter fits (symbols) for six observables.
}
\end{figure}

%
%\begin{eqnarray*}
%  p = (\vec{p},E) & \longrightarrow &(\vec{p},\left|\vec{p}\right|)\qquad\mbox{({$p$}-scheme)} \\
%  p = (\vec{p},E) & \longrightarrow &(\hat{p}E,E) \qquad\mbox{({$E$}-scheme)}
%\end{eqnarray*}
%
This redefinition doesn't influence the power correction term that is
calculated for massless particles. The difference between the schemes is
therefore an indicator for the size of the effect of hadron masses.
As shown in \fig{fig:masseffects} it can be of significant size.
Of the different schemes the $E$-scheme is singled out, 
because in this scheme the mass correction is
universal and can thus to a good approximation be absorbed into the standard
power correction~\cite{Salam:2001bd}.

Extra transverse momentum from the decay of hadrons, especially $B$-hadrons,
does similarly change the observables, but can't be significantly reduced 
by the redefined scheme. Their influence including the varying initial rate of
$b$-quarks need to be accounted for otherwise.

The DELPHI analysis presented in the following section takes the $B$-mass 
effects into account by applying a MC correction
and uses the $E$-scheme definition for jet masses to reduce the influence of
stable hadrons.

\section{RGI}
The renormalisation group invariant perturbation theory 
(RGI)\cite{Dhar:1984py} uses the
observable itself as expansion parameter. Thus it has no dependence on the
renormalisation scale. 

For a mean event shape $R=\langle f \rangle/A_f$ RGI connects the energy
dependence of an observable with the observables $\beta$-function:
\beq
Q\frac{\mathrm{d}R}{\mathrm{d}Q}
= \beta_R(R)=-\frac{{\beta_0}}{2\pi}R^2
 \left(1+\frac{\beta_1}{2\pi\beta_0}R+\rho_2R^2+\ldots\right)
\label{eq:rgi}
\eeq
As usual, this can be solved introducing an integration constant, $\Lambda_R$.
 $\Lambda_R$ can be connected to
$\Lambda^{\mathrm{QCD}}_{\overline{\mathrm{MS}}}$\cite{Celmaster:1979km}
and thus be used to measure the strong coupling.

% Unterschied zwischen beta Messung und
% alpha_s Messung besser Formulieren.

The \as-results  obtained by the DELPHI collaboration with this method
show good consistency of $\alpha_s=0.119\pm 0.004$ when including RGI power
corrections. (The quoted variation indicates the spread of 6
observables.) 
As the power corrections are consistent with zero, the data can
be described without power correction leading to an even improved consistency
of $\alpha_s=0.117\pm0.002$. 
That data can be described without power corrections or hadronisation
corrections raises the  question, whether power corrections which usually have
a size of around 10\% at $M_Z$ are to a large part perturbative. 

In fact RGI is able to 'predict' a value for \asb\ as function of \as,
from setting the RGI prediction equal to the power correction formula.
\fig{fig:a0prediction} shows \asb, as obtained after choosing 
the PDG average of $\alpha_s=0.118$, compared to those obtained with the
2-parameter power corrections fits. The 'predictions' (blue bands) agree with
the fit results (symbols) much better than any universal \asb-value around the
expected 0.5.

\section{Measurement of the $\beta$-function}

The RGI formula \eq{eq:rgi} 
allows to deduce the $\beta$-function of an observable
directly from the energy evolution of its measured mean values. Because the
$\beta$-functions are universal to NLO this is equivalent to a measurement of
\em the \em  $\beta$-function of QCD.

\begin{figure}[t]
  \begin{center}
\epsfig{file=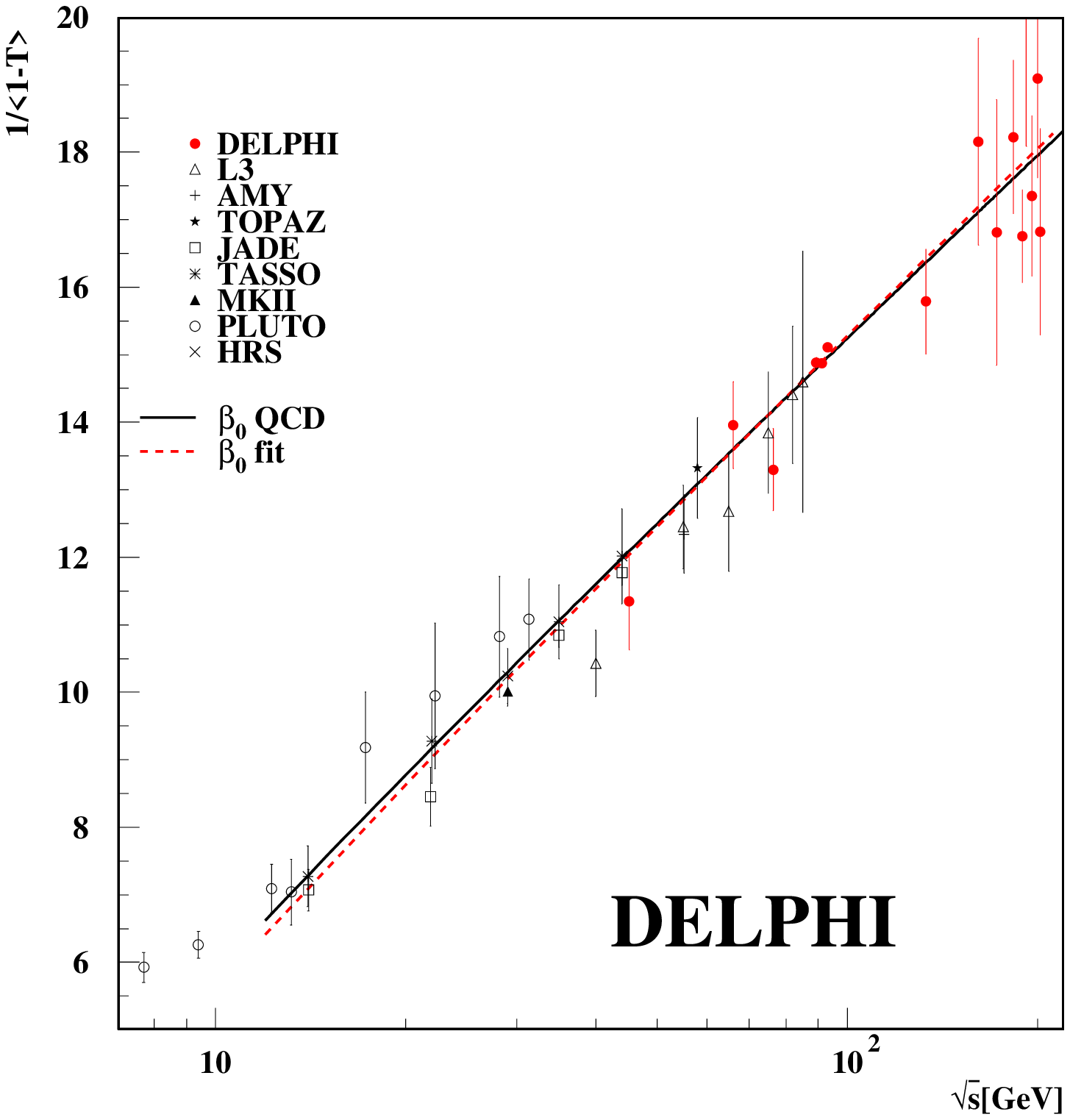,width=0.518\textwidth}
\epsfig{file=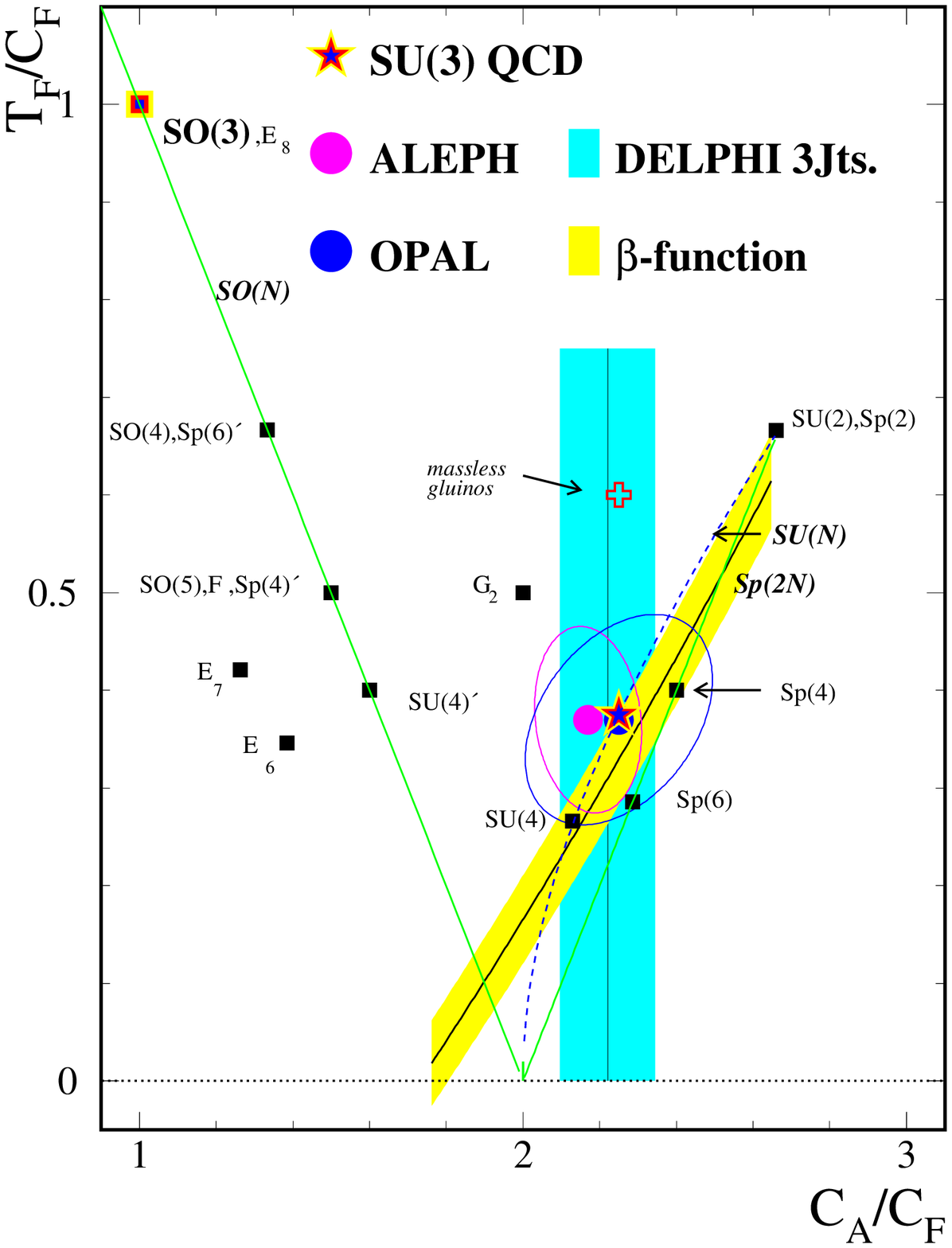,width=0.382\textwidth}
  \end{center}
\caption{\label{fig:beta}%
Measurement of the $\beta$-function using mean Thrust.
The lefthand-side shows measurement of the inverse of mean $1-T$ as function
of energy compared to a two parameter fit and to the QCD $\beta$-function.
The righthand-side compares the constraint put by this results on the
$T_R/C_F$ vs.\ $C_A/C_F$ plane to results of other measurements.
}
\end{figure}

Experimentally the energy dependence is measured as slope of the inverse mean
value as function of the logarithm of the energy, which measures 
$-R^2\beta_R(R)$.
%\beq
%\frac{\mathrm d R^{-1}}{\mathrm d \log Q}
%=\frac{\beta_0}{2\pi}\left(1+\frac{\beta_1}{2\pi\beta_0}R+\ldots\right)
%\eeq
From DELPHI thrust data one finds 
${\mathrm d R^{-1}}/{\mathrm d \log Q}=1.35\pm0.16$;
including data from low energy experiments the fit results in
${\mathrm d R^{-1}}/{\mathrm d \log Q}=1.38\pm0.05$.
\fig{fig:beta} (left) picturates the excellent data description of this fit
and the good agreement with the QCD expecation of 1.32. 
When interpreted within QCD these obtained slopes
correspond to a number of active flavours of $n_\mathrm{f}=4.7\pm0.7$ 
and $n_\mathrm{f}=4.75\pm0.44$,
respectively.

As the $\beta$-function is a function of the structure constants, it is
interesting to compare the constraints of this measurements with other
measurements of the structure constants, which are usually plotted in the
$n_fT_R/C_F$ vs.\ $C_A/C_F$ plane.
In \fig{fig:beta} (right)\cite{KH_Sienna2001} 
the result for the $\beta$-function
is  compared to results from angular distributions 
of ALEPH\cite{Bravo:2001ga} and OPAL\cite{Abbiendi:2001qn}, and
to a result from gluon jet multiplicity of DELPHI\cite{Hamacher:2000cn}. 

It shows that the structure is by now very well confirmed by measurements and
that it poses strict constraints on new physics, excluding e.g. light gluinos.

\section{Conclusions}
Event shapes %observables 
in $e^+e^-$ collisions directly probe properties of
hard QCD. At LEP they can be measured with a
precision that exceeds the precision of the current theoretical understanding.
The investigation of the consistency of \as-results 
from several such event shapes 
gives hints in which areas problems still hide 
or which renormalisation scheme may be singled out.

As (most of) the calculations don't take  mass effects
into account, the influence of the $B$-hadron mass
must be corrected for and the influence of stable hadron masses 
should be minimised
by using observables without explicit mass dependence.

Compared to the standard method (\oas+NLLA in $\log R$-scheme) the
RGI method shows a dramatically reduced spread in \as\ results.
This improved consistency awaits theoretical explanation.
RGI can further be used for a measurement of the $\beta$-function from mean 
thrust with a remarkable precision, excluding recently discussed light gluinos.
\section*{References}
\bibliography{Moriond,QCD}
\end{document}